# Tuning of dielectric properties and magnetism of SrTiO$_3$ by site-specific doping of Mn

D. Choudhury[1,2], S. Mukherjee,[1] P. Mandal[3], A. Sundaresan[3], U. V. Waghmare[4], S. Bhattacharjee[5], R. Mathieu[6], P. Lazor[7], O. Eriksson[5], B. Sanyal[5], P. Nordblad[6], A. Sharma[2], S. V. Bhat[2], O. Karis[5], D. D. Sarma[1,5,*]

[1] *Solid State and Structural Chemistry Unit, Indian Institute of Science, Bangalore-560 012, India.*

[2] *Department of Physics, Indian Institute of Science, Bangalore-560012, India.*

[3] *Chemistry and Physics of Materials Unit, Jawaharlal Nehru Centre for Advanced Scientific Research, Bangalore-560064, India.*

[4] *Theoretical Science Unit, Jawaharlal Nehru Centre for Advanced Scientific Research, Bangalore-560064, India.*

[5] *Department of Physics and Astronomy, Uppsala University, Box 516, SE-75120, Uppsala, Sweden.*

[6] *Department of Engineering Sciences, Uppsala University, SE-75121, Uppsala, Sweden.*

[7] *Department of Earth Sciences, Uppsala University, SE-75236, Uppsala, Sweden.*

Combining experiments with first principles calculations, we show that site-specific doping of Mn into SrTiO$_3$ has a decisive influence on the dielectric properties of these doped systems. We find that phonon contributions to the dielectric constant invariably decrease sharply on doping at any site. However, a sizable, random dipolar contribution only for Mn at the Sr site arises from a strong off-centric displacement of Mn in spite of Mn being in a non-$d^0$ state; this leads to a large dielectric constant at higher temperatures and gives rise to a relaxor ferroelectric behavior at lower temperatures. We also investigate magnetic properties in detail and critically reevaluate the possibility of a true multi-glass state in such systems.

I. INTRODUCTION

SrTiO$_3$ is an easily functionalizable material and has been the playground for investigations of a variety of phenomena [1]. By itself, it is a wide band-gap, diamagnetic and quantum paraelectric insulator that is used widely as a substrate material for growths of thin films and for fabrications of prototypical devices. In the presence of oxygen vacancies it can exhibit superconductivity [2], and metallic nature [3], as well as switchable resistive states [4] depending on the extent of doping. Growing a thin layer of another wide bandgap insulator, such as LaAlO$_3$, on top of SrTiO$_3$ leads to the realization of a high mobility two-dimensional electron gas at the interface [5]. SrTiO$_3$ can be doped with a variety of elements at the A-site (Sr site) and heterovalent dopants, such as La$^{3+}$ substituting Sr$^{2+}$, makes the system an n-type metal.

In this paper, we focus on substitution at both A and B sites with a transition metal ion, namely Mn. Mn not only can support a local magnetic moment, but also can exist in a wide variety of oxidation states, thereby giving rise to a possibility of doping charge carriers. With reference to yet another interesting physical property of SrTiO$_3$, namely its high dielectric constant and low loss ($\varepsilon'_r$ =248 and $D$ = 0.002 at 1 MHz and 300 K), doping at the B site may *a priori* be anticipated to have drastic effects, since the dielectric properties are intimately connected with the $3d^0$ state of Ti$^{4+}$ at the B site. Interestingly, however, past experimental results indicate a more drastic effect on dielectric properties arising from A-site substitution rather than the B-site substitution. For example, Mn substitution at the Ti site of SrTiO$_3$ (TiMn ceramics) remains paraelectric [6,7] as SrTiO$_3$ [8], while Mn substitution at the Sr site (SrMn ceramics) gives rise to qualitatively different, relaxor ferroelectric properties at temperatures (T) below 150 K [6, 9].

It is well-known that random substitutions as well as other kinds of disorder may have profound effects on the electronic and magnetic properties of transition metal oxides [10,11]. However, associated with the issue of such doping of transition metal ions with local magnetic moments is the question whether SrTiO$_3$ can also be rendered magnetic. This would open up the exciting prospect of a multiferroic state, considering the proximity of SrTiO$_3$ to a ferroelectric instability. There have indeed been recent reports

[12,13], suggesting the coexistence of a glassy magnetic state and a glassy dielectric behavior in certain Mn-doped SrTiO$_3$ system, defining a new class of systems, termed "multi-glass" in analogy to multiferroics.

In the present work, we dope Mn independently at Sr and Ti sites, thereby producing, not only the known SrMn and TiMn compounds, but also a novel system (SrMnTiMn) with Mn simultaneously substituting at both Sr and Ti sites, helping us to critically probe the origins of such a diverse panorama of properties. While there have been limited efforts to understand the origin of such unusual dielectric behaviors with the help of *ab initio* studies [14,15], these studies were necessarily limited in their scopes due to the enormous complexity of the system involving dilute substitution and phonon calculations. We use state of the art *ab initio* results to understand the observed properties in detail. In addition, we critically evaluate the recent claim of a multi-glass state in such systems [12, 13], pointing to the possibility of this being an extrinsic phenomenon.

## II. METHODS

Mn doped SrTiO$_3$ samples were synthesized with Mn ions controllably substituting only the Sr site, Sr$_{0.98}$Mn$_{0.02}$TiO$_3$ (SrMn), only the Ti site, SrTi$_{0.9}$Mn$_{0.1}$O$_3$ (TiMn) or both Sr and Ti sites, Sr$_{0.97}$Mn$_{0.03}$Ti$_{0.93}$Mn$_{0.07}$O$_3$ (SrMnTiMn), following the conventional mixed oxide method [6, 12], starting with SrCO$_3$, TiO$_2$ and Mn$_2$O$_3$ in stoichiometric amounts. Electron Paramagnetic Resonance (EPR) experiments were performed with an ER 200D X-band Bruker Spectrometer at 9.8 GHz. Room temperature x-ray absorption (XA) measurements at the Mn $L_{2,3}$ edge were performed using linearly polarized synchrotron light at I1011 beamline, MAXlab, Sweden. The XA spectrum was recorded using the total electron yield method by recording the sample drain-current as a function of the incident photon energy. Dielectric measurements were performed using an Agilent 4284A LCR meter in the temperature range from 10 K to 300 K. Dielectric constant values, obtained separately with sputtered gold or silver paste as electrodes, were found to be same, thereby ruling out any electrode polarization contribution. DC magnetization and AC susceptibility of the samples were recorded as a function of the temperature and the magnetic field using a superconducting quantum interference device magnetometer from Quantum Design. Specific heat measurements were performed using a physical property measurement system from Quantum design. Room temperature unpolarized micro-Raman spectra were collected in the back-scattering geometry, in the range 150-2340 cm$^{-1}$, at a resolution of ~3 cm$^{-1}$ using an Argon-ion laser [16].

First-principles calculations were performed within density-functional theory, as implemented in the Vienna ab-initio simulation package (VASP), with local spin density approximation (LSDA), projector augmented-wave (PAW) potentials [17] and Hubbard U-corrections for the *d*-states of Mn. We used a 2x2x2 supercell to simulate 12.5% Mn concentration. For these calculations, we optimized the crystal structures of doped as well as undoped samples with respect to variations in the lattice vectors as well as the coordinates of all atoms within the 2x2x2 supercell by minimizing the Hellman-Feynman forces. However, this structural optimization does not capture the antiferrodistortive motions of the oxygen octahedra [18]. In each case, we estimated the dielectric constant with contributions from electrons and IR-active phonons [19] for the geometry optimized crystal structures. The Born effective charges were calculated using density functional perturbation theory [20] and the phonon frequencies were determined by frozen phonon calculations using the supercell. The dielectric constants were calculated following the method described by Cockayne and Burton [21].

## III. RESULTS AND DISCUSSIONS

Room temperature X-ray diffraction of all Mn-doped SrTiO$_3$ samples (SrMn, TiMn and SrMnTiMn) were found to crystallize in cubic *Pm-3m* space group. The unit cells were observed to decrease for all Mn-doped samples, as evidenced by the shift of the 2θ peaks in the x-ray diffraction pattern towards higher angles for doped samples. Also, reductions in lattice parameters, caused by the doping of smaller Mn ions into the SrTiO$_3$ lattice, were found to be similar to reported values [6]. The reduction in lattice parameters of SrMnTiMn compared to SrTiO$_3$ agrees with an algebraic sum of observed unit cell reductions brought about by individual site dopings [6], according to the expected stoichiometry. Electron paramagnetic resonance (EPR) spectroscopy, well established in elucidating local environment as well as the valency of doped Mn ions in SrTiO$_3$ lattice [7,22,23], was used to characterize present samples. The EPR spectrum, recorded on SrMn, shown in Fig. 1, consists of a sextet of hyperfine split lines, with Lande $g$ factor of 2.01 and a hyperfine splitting of 88 Oe, being characteristic of Mn$^{2+}$ ions doped at Sr site in SrTiO$_3$ [23]. The EPR spectrum of the sample with Mn at the Ti site also consists of a single sextet of hyperfine split lines (Fig. 1), with Lande $g$ factor of 1.997 and a hyperfine splitting of 75 Oe, being characteristic of Mn$^{4+}$ ions doped at the Ti site in SrTiO$_3$ [7, 22]. The EPR spectrum of SrMnTiMn exhibits an overall broad signal, well-known to arise at such higher concentrations of Mn doping due to dipolar interactions. However, the same spectrum also clearly shows the existence of finer features arising from hyperfine interactions. There is an obvious one-to-one correspondence between hyperfine features from SrMnTiMn and those from SrMn and TiMn, as indicated by vertical lines in Fig-1, thus establishing the presence of two species of Mn ions in SrMnTiMn; one in the form of Mn$^{2+}$ ions doped at the Sr site and the other in the form of Mn$^{4+}$ ions at the Ti site of SrTiO$_3$. XA spectroscopy, known to provide very distinct spectral shapes for different valencies of Mn ions [24], was used to further distinguish Mn$^{2+}$ dopants at Sr site from Mn$^{4+}$ dopants at the Ti site. The characteristic features of Mn $L_{II,III}$ XA spectra, in comparison with reference spectra, as shown in Fig. 2, clearly establish the Mn$^{2+}$ valence state in SrMn and Mn$^{4+}$ valence state in TiMn samples. Thus, XAS results, indeed confirm the EPR findings. In order to bring out clearly the difference in EPR spectra between Mn$^{2+}$ ions doped at the Sr site from Mn$^{2+}$ ions doped at the Ti site of SrTiO$_3$, we have prepared a sample with Mn$^{2+}$ at the Ti site by careful reduction of the sample starting from Mn$^{4+}$ at the Ti site. The EPR spectrum of this reduced sample has a single sextet of lines with a hyperfine splitting of 83.5 Oe, being distinctly different from the EPR spectra of SrMn sample, which has a hyperfine splitting of 88 Oe, clearly bringing out the specificity of EPR to the crystallographic site of doping.

We find that Mn doping at the Ti site retains the quantum paraelectric state and does not have any qualitative influence on dielectric properties in comparison to pure SrTiO$_3$, as reported earlier [25]. Quantitatively, the dielectric constant ($\varepsilon'_r$) is less, while the loss ($D$) is distinctly higher in the case of the doped sample, as shown in Table I. In contrast to doping at the Ti site, $\varepsilon'_r$ and $D$ of SrMn ceramic, measured as a function of the temperature for various frequencies show qualitatively different behaviors, as shown in Fig. 3(a). The dielectric constants of SrMn exhibit a strong frequency-dependent relaxation, with the temperature ($T_m$) for the maximum dielectric constant relaxing with the corresponding measurement frequency ($f_i$). If we apply the same critical slowing down analysis, usually employed to study the second-order spin glass phase transition [26], to the variation of $T_m$ with frequency $f_i$, we observe that the dielectric relaxation follows the power law $f_i \propto f_o \times (T_m/T_{gl} - 1)^{zv}$, with the glass transition temperature ($T_{gl}$), critical exponent ($zv$) and $f_o$ being 40 K, 10.6 and $1.3 \times 10^6$ Hz, respectively. Thus, the present SrMn ceramic behaves like a dielectric glass, similar to previous results [12, 13]. The dielectric glassy behavior is expected to originate from the off-centric position of small $Mn^{2+}$ ion ($r_A$=1.25 Å) at the $Sr^{2+}$ ($r_A$=1.44 Å) site, as suggested by extended x-ray absorption fine structure (EXAFS) studies [27], thereby giving rise to multiple ground states [14, 27]. Interestingly, Table I shows a clearly contrasting quantitative behavior between dopings at Ti and Sr sites. Instead of the sharp decrease in case of TiMn, $\varepsilon'_r$ of SrMn in fact shows a modest increase, with the dielectric loss retaining a favorable value (Table I).

Dielectric properties of the sample with Mn-doped at both Sr and Ti sites (Fig. 3(b)) are qualitatively similar to those of SrMn sample (Fig. 3(a)), exhibiting glassy dielectric properties below 100 K, as suggested by the dielectric relaxation as a function of the frequency. The dielectric relaxation for SrMnTiMn obeys the same power law relation as that of SrMn, with values of $T_{gl}$, $zv$ and $f_o$ being 35 K, 7.8 and $9 \times 10^7$ Hz, respectively. These values are very similar to those for the previously discussed SrMn sample, discussed above, establishing this as a dielectric glass at low temperatures arising from off-centric movements of Mn doped at the Sr site. More interestingly, we find that the dielectric loss in this new compound is better by an order of magnitude (Table I), while still retaining the high dielectric constant value of $SrTiO_3$. This shows that the co-doping at the Ti site allows us to tune the loss, while largely retaining the advantages of a high dielectric constant as in the case of Mn doping at the Sr site.

Motivated by recent claims of observation of intrinsic glassy magnetic properties in SrMn ceramic [12,13] and in a related $K_{0.97}Mn_{0.03}TaO_3$ system [28], we have performed magnetic studies of various Mn doped $SrTiO_3$ systems. In contrast to these earlier claims, we do not observe any indication of a magnetic ordering of any type in *M(H)* (see inset to Fig. 3a for SrMn) and *M(T)* measurements (not shown here) down to the lowest temperature of about 10 K for SrMn and TiMn samples. Instead we observe linear dependencies of *M* vs. *H at T*=10 K and Curie-Weiss like behavior of the susceptibility. However, it turns out that magnetic properties of our new SrMnTiMn sample (Fig. 4(a)) are remarkably similar to behaviors of the SrMn sample reported in Ref. 12 and 13, suggesting it to be weakly magnetic, with a broad magnetic response below 60 K and a frequency dependence of the a.c. susceptibility (inset II to Fig. 4(a)). The glassy nature of the observed magnetism, was further confirmed by memory experiments performed by recording differences of zero field cooled (ZFC) magnetization measurements both as a function of different wait temperatures as well as different wait times [12]. Results of such memory experiments on the SrMnTiMn compound (Fig. 4(b)), clearly show sharp dips at the respective wait temperatures, whose depths increase with increasing wait times, clearly demonstrating the aging and memory features associated with glassy magnetic behavior. Specific heat data (Inset to Fig. 5) exhibit a broad component, which extends from around 70 K to 30 K, without any indication of a sharp magnetic transition at any given temperature. The fact that such magnetic signatures appear only for the SrMnTiMn sample, but not for the SrMn sample, while both exhibiting very similar glassy dielectric behavior suggests that glassy magnetic properties of the SrMnTiMn sample are not dependent on or associated with the glassy dielectric properties, though these two glassy states set in at approximately the same temperature (T $\leq$ 60 K). The decoupling of these two properties is even more directly evidenced by the observation of a complete absence of magneto-capacitance at all frequencies and temperatures, for SrMnTiMn shown in Fig. 5 with open symbols and lines representing dielectric constant for an applied field, *H*=0 and 1 T, respectively. It is significant that the present SrMn sample, while exhibiting a glassy dielectric behavior as reported before, did not show any magnetic anomaly in sharp contrast to earlier results [12, 13] on the same compound. This suggests that the observed weak magnetic response in SrMn compound of Refs. 12 and 13 and our SrMnTiMn sample to have an extrinsic origin.

In order to probe the possibility of an impurity phase in the sample, micro Raman spectroscopic measurements were performed on SrMnTiMn. In addition to the bulk Raman spectra, shown by the dashed black line in Fig 6, statistically one out of 100 spots on the SrMnTiMn sample showed Raman spectra [red (lower) line in Fig. 6] with additional peaks at 324, 373 and 660 cm$^{-1}$, which are characteristic lines of $Mn_3O_4$ in hausmannite phase [29]. XRD data with an exceptionally high signal-to-noise ratio plotted in the logarithmic scale (Inset to Fig. 6), reveals a broad impurity peak in the SrMnTiMn sample at about $2\theta = 36^o$, again characteristic of the presence of small clusters of Hausmanite $Mn_3O_4$. A Scherrer fit to the broad impurity XRD peak, yields an average size of 17 nm for these $Mn_3O_4$ clusters. In order to investigate this point further; we prepared several SrMnTiMn samples with the same target composition, namely $Sr_{0.97}Mn_{0.03}Ti_{0.93}Mn_{0.07}O_3$. Since the impurity formation is not a controllable process, we expected the impurity contribution to vary randomly from sample to sample. In every case we find the impurity level to be very small, as shown by the tiny impurity peak observable only in the logarithmic scale of XRD (see Fig.7inset). The main frame of the figure shows the corresponding magnetic data. The one to one correspondence of the observed magnetization values of different SrMnTiMn samples with the extent of the $Mn_3O_4$ impurity phase, as shown in Fig. 7, confirms the extrinsic origin of magnetization in SrMnTiMn sample. Since such impurity phases, present in tiny quantities, are difficult to characterize in detail, particularly with respect to its exact oxygen stoichiometry, we ascribe such extrinsic glassy magnetic behaviors to the presence of $Mn_xO_y$ phases with $x\sim3$ and $y\sim4$.

We now turn to the phonon spectra of these compounds in order to understand the microscopic origin of the observed trends in dielectric responses in doped $SrTiO_3$ compared to the undoped one. In Fig. 8, we present the phonon density of states (DOS) at the Gamma point for all undoped and doped systems calculated with the 2x2x2 supercell. It is clear that all compounds, including $SrTiO_3$, show a few unstable modes, appearing with imaginary frequencies in the DOS. This is due to the fact that the structural optimizations were performed without considering the antiferrodistortive rotational motion of oxygen octahedra [18]. Therefore the unstable modes due to such rotations are retained in the calculations and appear as modes with imaginary frequency in the calculated DOS. As can be seen from Fig. 8, these are non-polar modes and therefore, do not contribute to the dielectric constants of any of the compounds. We find that the structure undergoes negligible local distortions when Mn is doped at Ti site, as expected from its non-$d^0$ state, though optimized lattice parameters indicate an overall contraction in agreement with the experiment, leading to a hardening of phonons (Fig. 8). Since the host lattice also does not exhibit any off-centering of Ti, TiMn and $SrTiO_3$ are structurally very similar, with a clear reflection in their vibrational spectra (Fig. 8). Due to broken symmetry, the IR-active phonons of pure $SrTiO_3$ mix with other phonons in the case of TiMn, resulting in slightly broadened peaks of IR-active modes. As a net result, the phonon contribution to the dielectric response is reduced with Mn-substitution at the Ti-site. The phonon contribution to the dielectric constant for SrMn is even lower than that of TiMn, essentially due to an overall decrease of Born effective charges at Ti sites for SrMn compared to TiMn, in addition to a phonon hardening compared to pure $SrTiO_3$ similar to the case of TiMn. Interestingly, Mn-substitution at the Sr site leads to a rather large off-centering of Mn due to (a) its much smaller ionic size compared to Sr ions and (b) the fact that Mn does not normally favor 12-fold coordination. The local dipole moment induced by Mn off-centered from the Sr site is 3.9 Debye, in contrast to 0.03 Debye when doped at the Ti site. It appears that this large dipolar contribution more than overcomes the decrease in the phonon contribution for SrMn, in contrast to the case of TiMn, when compared with undoped $SrTiO_3$ as suggested by our experiments (Table I). It is interesting to note that the emergence of the relaxor behavior through the substitution of hetero-ferroactive substitution at the *A* site (Mn at the Sr site) is intriguing. We trace the origin of this behavior with the help of the vibrational spectrum. We find a rather strong mixing between modes of $SrTiO_3$ upon Mn substitution at Sr site, as reflected in broad and numerous IR-active modes in SrMn. Thus, a strong-mode coupling appears to be responsible for the relaxor behavior of SrMn. Finally, dielectric behavior of SrMnTiMn can be readily understood from a combination of factors discussed here for substitutions at Ti and Sr sites.

In order to investigate the effect of the on-site Coulomb interaction strength,$U$, on the dielectric properties of SrMn, dielectric constants, Born effective charges, and phonon frequencies were calculated for three different values of $U$ ($U$ = 2, 4 and 6 eV)

(see Table II). While it is found that the electronic contribution to the dielectric constant ($\varepsilon_\infty$) decreases with an increase in the value of $U$ due to a steady increase in the band gap, the corresponding lattice contribution to the dielectric constant ($\varepsilon_L$), however, increases due to enhanced Born effective charges, coupled with a softening of the softest phonon mode. We note here that the soft phonon mode softens by about 10 cm$^{-1}$ on going from $U = 2$ eV to $U = 6$ eV.

In order to investigate the relaxation behavior of samples with Mn doped at the Sr site, we studied the transition states using the nudged elastic band method. When Mn is doped at the Sr site, it is in a cuboctahedral co-ordination having 12 bonds with 12 oxygen atoms and 14 faces, as shown in Fig. 9(a). In the distorted cuboctahedron, obtained after relaxation of the geometry, as shown in Fig. 9(b), the Mn ion becomes off-centric, as also discussed earlier. The Mn ion can go off-centric either by displacing towards the centre of a bond between two oxygen atoms of the cuboctahedron (bond off-centering) or it can displace towards the centre of the triangular face formed by three oxygen ions (face off-centering). In our relaxed distorted cuboctahedron, we find that Mn ions preferentially off-centers along the centre of the bond joining two oxygen atoms (see Fig. 9(b)). We have computed the transition states for Mn in the distorted cuboctahedron for both these possible off-centering geometries, and find that the transition state energy for the bond-bond transition is about 100 meV, in good agreement with the activation energies (~90 meV), estimated for Mn ion relaxation from our dielectric data. It is interesting to note here that the energy barrier for reorientation of the impurity-induced dipole (~ 80 meV) in Li-doped KTaO$_3$ is close [31] to what is found here for Mn-induced dipole reorientation (~90 meV experimentally and ~100 meV theoretically); this also explains the similarity of the present dielectric properties to those of Li-doped KTaO$_3$. In contrast, the transition state energy for the face-face transition is about 600 meV, *i.e* approximately six times higher than what is observed for a bond-bond transition. This further shows that Mn ions energetically favours the bond off-centering, in good agreement with results obtained from geometry optimization.

## IV. SUMMARY

We have studied dielectric properties of various Mn doped $SrTiO_3$ systems, establishing that the dielectric constant and the loss can be significantly tuned by controlled doping at the Sr and Ti sites. Specifically, Mn-doping at the Ti site was found to leave the sample in the paraelectric state as in the case of the undoped sample, while reducing the dielectric constant drastically. In contrast, Mn-doping at the Sr site (with or without a co-doping of Mn at the Ti site) was found to alter dielectric properties by establishing a relaxor behavior at lower temperatures, while the near room temperature dielectric constant remains comparable to, or even slightly higher than, that of $SrTiO_3$. *Ab-initio* electronic structure and phonon spectral calculations provide a qualitative framework to understand these changes. Phonon hardening on substitution, indicated by the lattice contractions and confirmed by first-principles calculations, leads to a sharp drop in the phonon contribution to the dielectric constant. However, a remarkable off-centering of Mn only at the Sr site, unusual for a non-$d^0$ ion, gives rise to a large dipolar contribution to the dielectric constant, regaining a high dielectric constant at high temperatures and a relaxor ferroelectric behavior at lower temperatures. In general, our samples did not show any magnetic ordering or anomaly, in contrast to previous reports, except for the case of Mn simultaneously doped at both Ti and Sr sites. In this latter case, which exhibits a glassy magnetic state below about 50 K, we have observed a clear decoupling of dielectric and magnetic properties, which suggests the absence of a true "multiglass" state in these systems. In our SrMnTiMn sample showing both dielectric and magnetic glassiness, the glassy magnetic properties were found to have an extrinsic origin, arising from $Mn_3O_4$ impurities in tiny quantities.

## V. ACKNOWLEDGEMENTS

The authors thank the Department of Science and Technology and the Board of Research in Nuclear Sciences, Government of India, the Swedish Foundation for International Cooperation in Research and Higher Education (STINT), the Swedish Research Council (VR), Swedish Research Links programme under SIDA, and the

European Research Council (ERC) for funding. O.E. also acknowledges ERC (project 247062 - ASD) and the KAW foundation.

**References:**

*Also at Jawaharlal Nehru Center for Advanced Scientific Research, Bangalore-560054, India and CSIR-Network of Institutes for Solar Energy.
Electronic mail: sarma@sscu.iisc.ernet.in

**Table I:** A comparison of experimentally obtained dielectric values of various Mn doped SrTiO$_3$ at 300 K and 1 MHz.

|          | SrTiO$_3$ | TiMn | SrMn  | SrMnTiMn |
|----------|-----------|------|-------|----------|
| $\varepsilon_r'$ | 248       | 140  | 278   | 251      |
| D        | 0.002     | 0.03 | 0.005 | 0.0003   |

**Table II:** Electronic ($\varepsilon_\infty$) and ionic ($\varepsilon_L$) contributions to the dielectric constant of SrMn for various values of $U$, obtained from our calculations.

| $U$ (eV) | $\varepsilon_\infty$ | $\varepsilon_L$ |
|----------|----------------------|-----------------|
| 2        | 6.97                 | 61.96           |
| 4        | 6.57                 | 65.43           |
| 6        | 6.49                 | 68.86           |

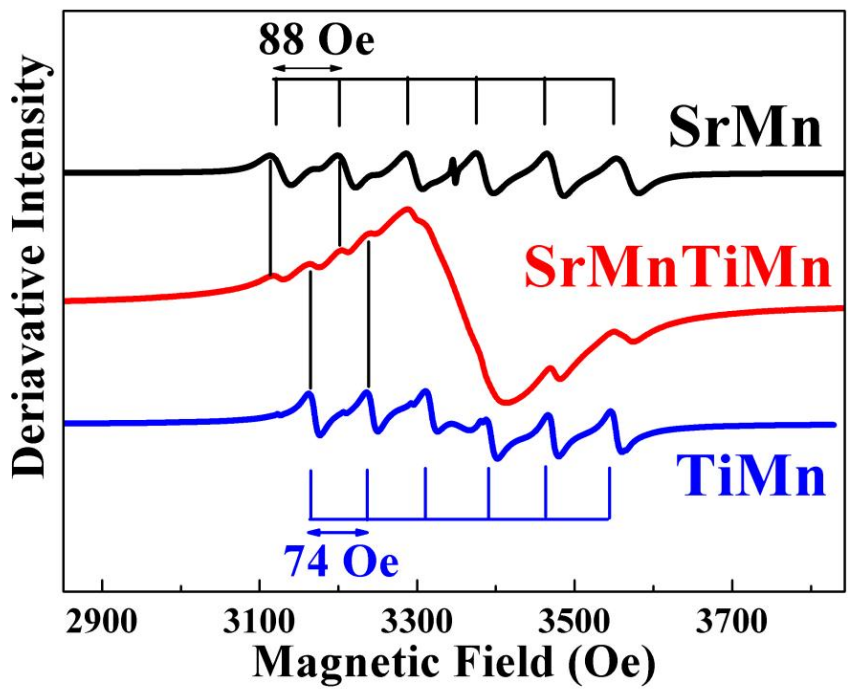

**Fig. 1**: (Color Online) Electron paramagnetic resonance (EPR) spectra of SrMn, SrMnTiMn and TiMn. To probe the Mn state at the Ti site, we specifically prepared a low doping TiMn sample with the composition $SrTi_{0.99}Mn_{0.01}O_3$ to avoid any dipolar broadening due to Mn-Mn interaction, as can be seen at higher concentrations of Mn, *e.g.* for the SrMnTiMn sample.

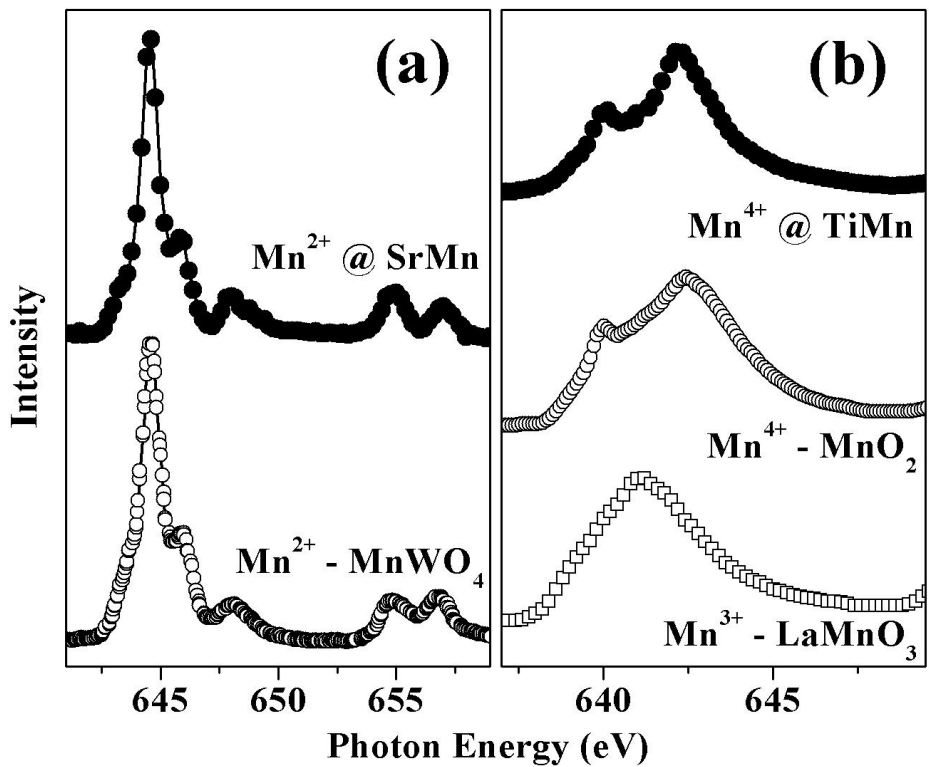

**Fig. 2**: Mn $L_{II, III}$ x-ray absorption spectra of SrMn and TiMn samples along with reference compounds, showing the presence of (a) $Mn^{2+}$ ions in SrMn sample and (b) $Mn^{4+}$ ions in TiMn sample.

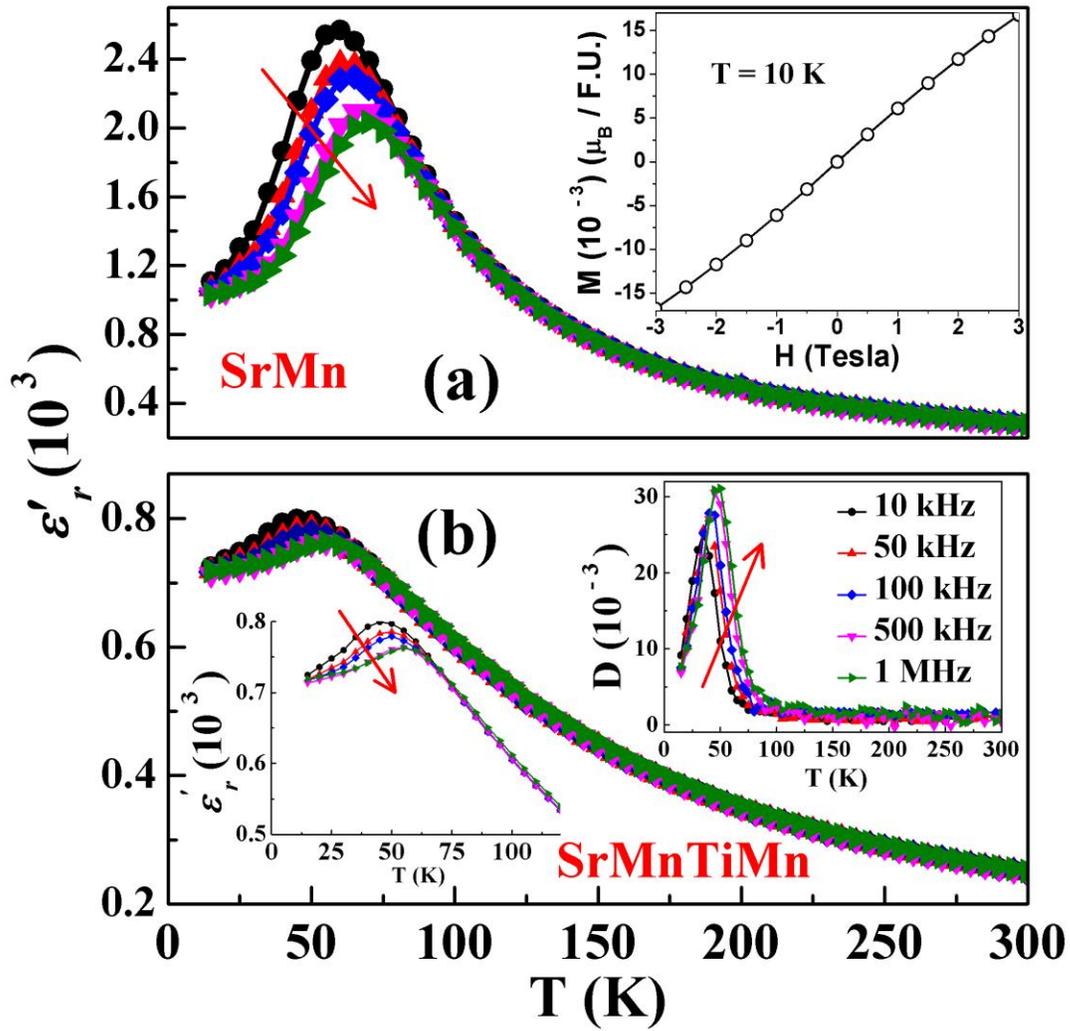

**Fig. 3:** (Color Online) (a) Dielectric constants of SrMn. Inset shows linear M-H measured at 10 K of SrMn. (b) Dielectric constants of SrMnTiMn for various fixed frequencies. Inset shows the corresponding loss (*D*). The arrow indicates the direction of increasing frequency in every panel.

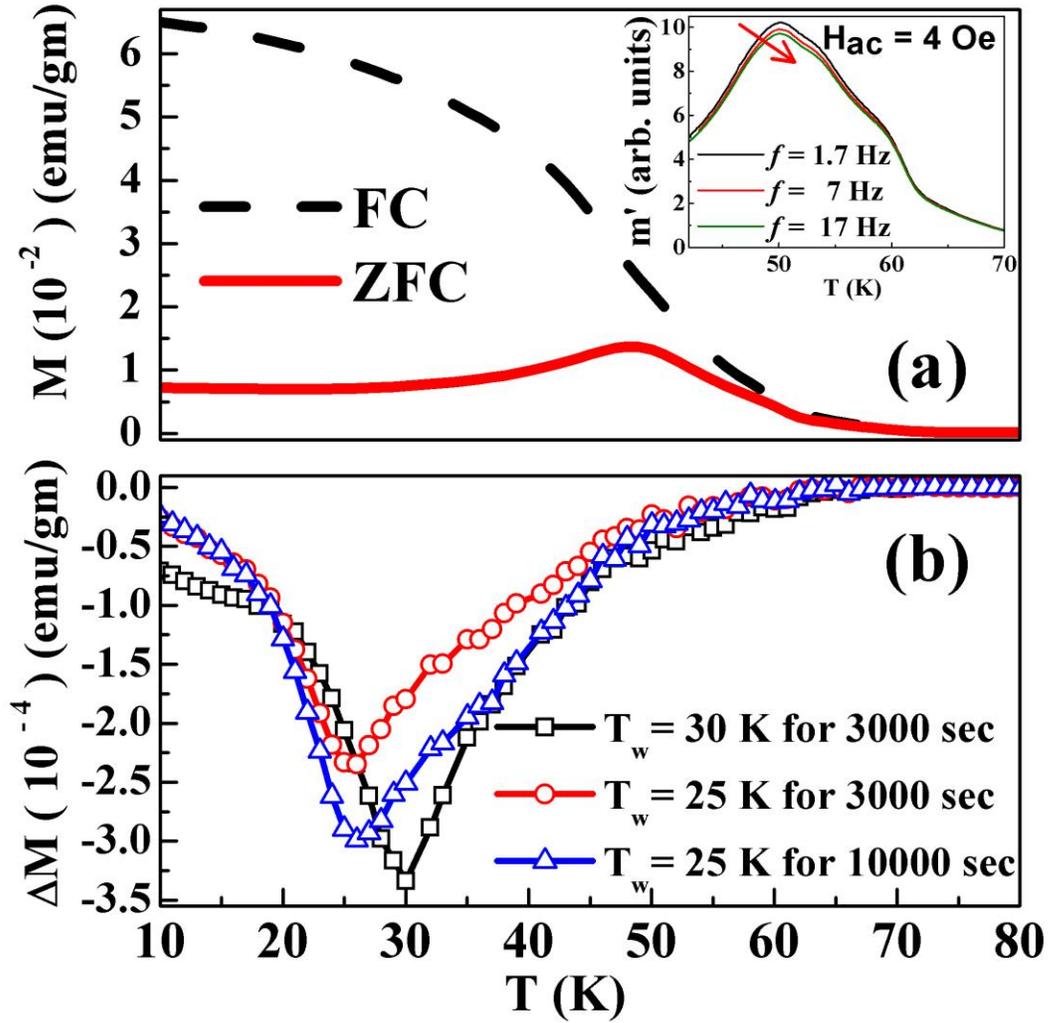

**Fig. 4**: (Color Online) Properties of $Sr_{0.97}Mn_{0.03}Ti_{0.93}Mn_{0.07}O_3$ (a) The field cooled (FC) and zero-field cooled (ZFC) d.c. magnetization measured with an applied d.c. magnetic field of 100 Oe with the inset showing the real part of the a.c. susceptibility data measured for an applied a.c. magnetic field of 4 Oe for a few frequencies. The arrow indicates the direction of increasing frequency. (b) The difference in ZFC data with intermittent waits ($T_w$) at 25 K and 30 K for different times as indicated in the plots.

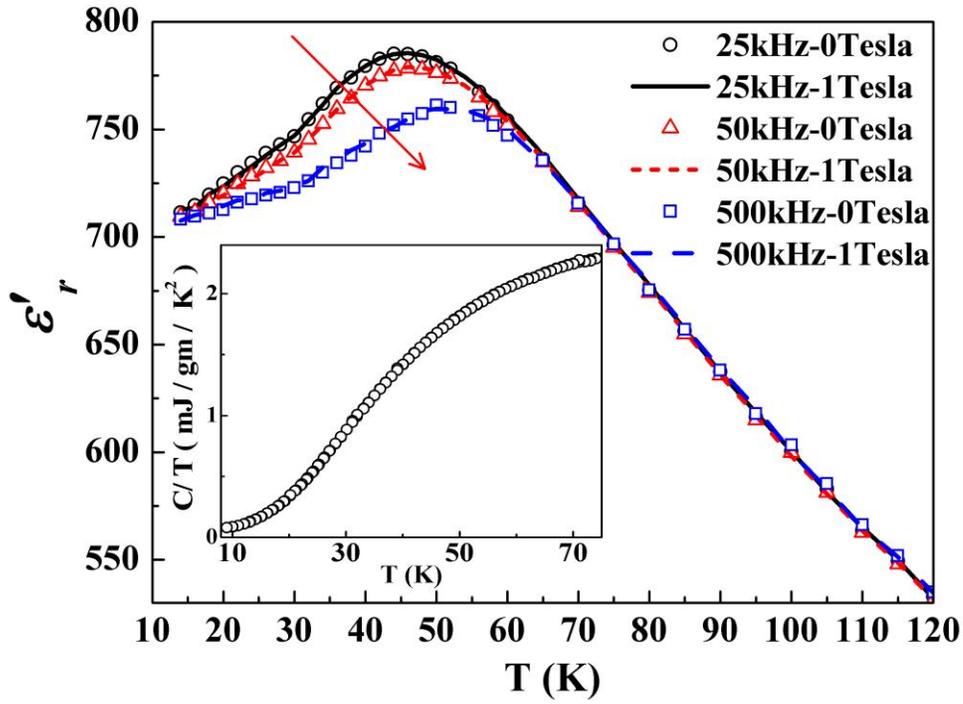

**Fig 5**: (Color Online) Dielectric constant of $Sr_{0.97}Mn_{0.03}Ti_{0.93}Mn_{0.07}O_3$ as a function of temperature for various fixed frequencies in the absence (circles) and in the presence (lines) of a magnetic field of 1 Tesla. The inset shows the specific heat as a function of temperature. The arrow indicates the direction of increasing frequency.

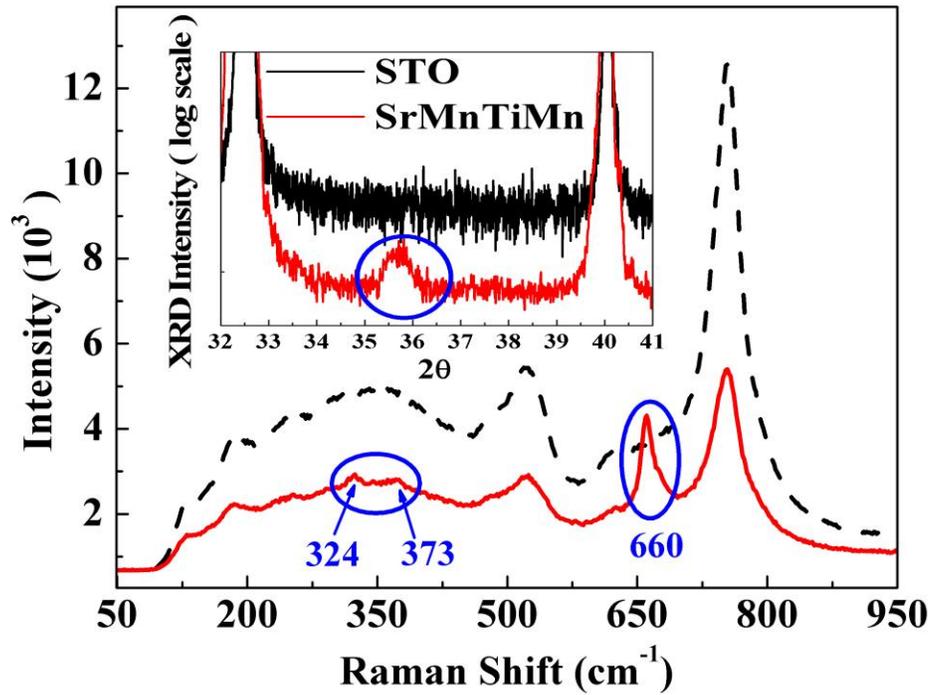

**Fig 6:** (Color Online) The bulk Raman spectra of SrMnTiMn as shown by the black dashed line. The red solid line shows the Raman spectra from the impurity phase overlapped on top of the bulk signal, which consists of characteristic lines of $Mn_3O_4$, highlighted by blue colored circles. The inset shows a detailed XRD scan.

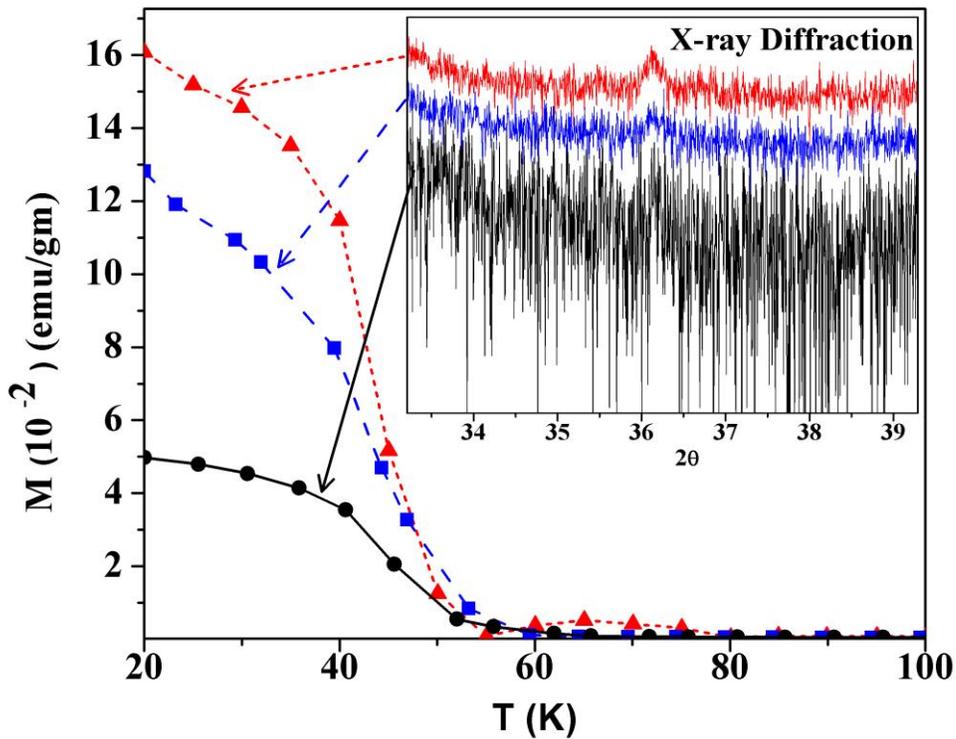

**Fig 7**: (Color Online) Field-cooled magnetizations (M) of various SrMnTiMn samples measured with an applied d.c. magnetic field of 100 Oe, with the corresponding x-ray diffraction patterns (shown in the inset), focusing on the 100% peak position of hausmannite $Mn_3O_4$ ($2\theta = 36^o$) [30], showing correspondingly varying amounts of $Mn_3O_4$ impurity phase in these samples.

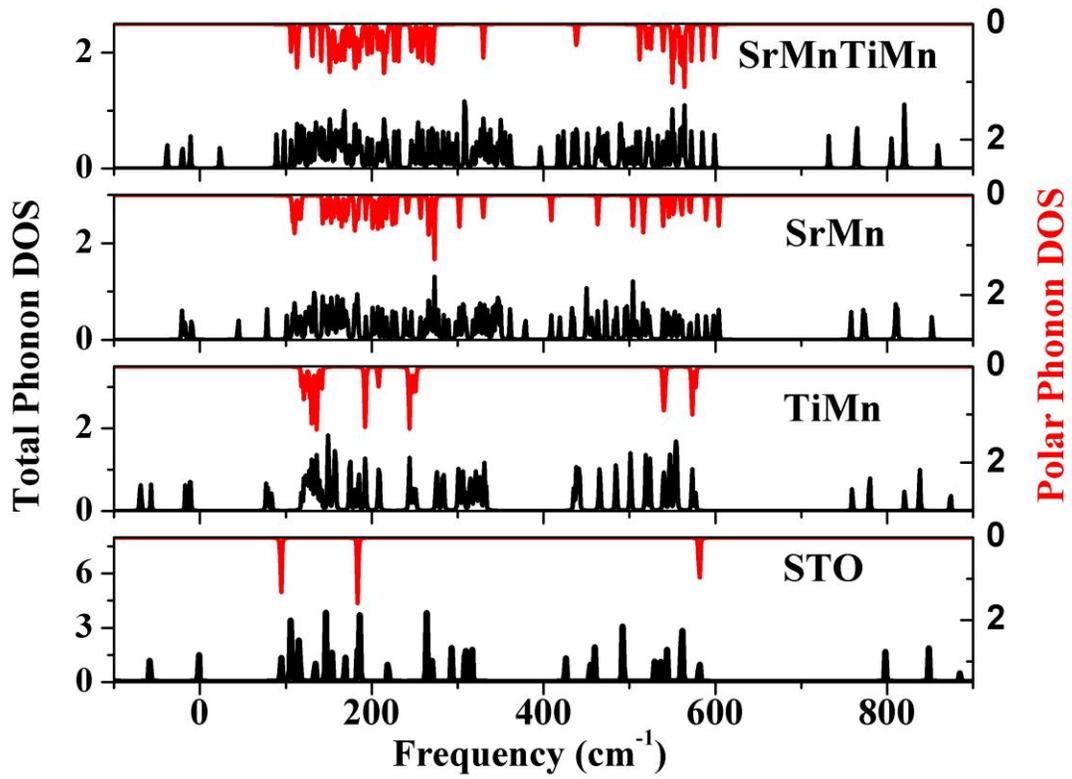

**Fig. 8**: (Color Online) Calculated total and polar phonon density of states at the Γ point (*q*=0) for various compounds as indicated. Plots are normalized to have the same area in every case.

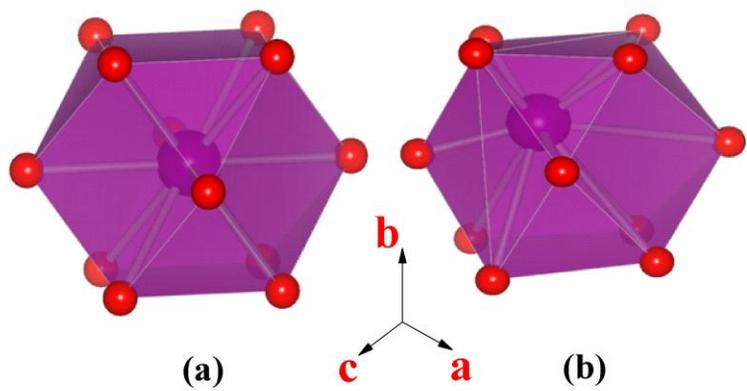

**Fig. 9**: (Color Online) Undistorted (a) and geometry optimized distorted (b) cuboctahedron of Mn ion in Mn doped at the Sr site of $SrTiO_3$. Mn and O atoms are drawn as purple (large) and red (small) balls respectively.